\documentclass[letterpaper, 10 pt, conference]{cssconf}
\IEEEoverridecommandlockouts             
\usepackage{cite}
\usepackage{graphicx}
\usepackage{amsmath}
\usepackage{amssymb}
\usepackage{siunitx}
\usepackage{xcolor}
\usepackage{tikz}
\usepackage{silence}
\usepackage{breqn}

\newtheorem{defn}{Definition}
\newtheorem{remark}{Remark}
\newtheorem{theorem}{Theorem}
\newtheorem{assumption}{Assumption}
\usetikzlibrary{arrows.meta, decorations.pathreplacing, arrows}
\DeclareMathOperator{\proj}{Proj}
\DeclareMathOperator{\spanop}{span}
\DeclareMathOperator{\inter}{int}
\NewDocumentCommand{\qfrac}{smm}{%
  \dfrac{\IfBooleanT{#1}{\vphantom{\big|}}#2}{\mathstrut #3}%
}

\title{\LARGE \bf
Nonlinear MPC for Tracking for a Class of Non-Convex Admissible Output Sets\thanks{A preliminary version of this paper was presented in \cite{cotorruelo2018tracking}.}
}

\author{Andres Cotorruelo, Daniel R. Ramirez, Daniel Limon, Emanuele Garone
\thanks{This research has been funded by the FNRS MIS “Optimization-free  Control  of  Nonlinear  Systems  subject  to  Constraints”,  Ref.F.4526.17.   by  Ministerio  de  Econom\'ia  y  Competitividad  of Spain under project DPI2016-76493-C3-1-R and by Ministerio de Ciencia e Innovación of Spain under project PID2019-106212RB-C41.}
\thanks{A. Cotorruelo and E. Garone are with Service d'Automatique et d'Analyse des Syst\`emes, Universit\'e Libre de Bruxelles, Belgium {\tt\small \{acotorru, egarone\}@ulb.ac.be}}%
\thanks{D. R. Ramirez and D. Limon are with Departamento de Sistemas y Autom\'atica, Universidad de Sevilla, Spain
        {\tt\small \{danirr, dlm\}@us.es}}%
}

\begin{document}

\maketitle
\thispagestyle{empty}
\pagestyle{empty}

\begin{abstract}
This paper presents an extension to the nonlinear Model Predictive Control for Tracking scheme able to guarantee convergence even in cases of non-convex output admissible sets. This is achieved by incorporating a convexifying homeomorphism in the optimization problem, allowing it to be solved in the convex space. A novel class of non-convex sets is also defined for which a systematic procedure to construct a convexifying homeomorphism is provided. This homeomorphism is then embedded in the Model Predictive Control optimization problem in such a way that the homeomorphism is no longer required in closed form. Finally, the effectiveness of the proposed method is showcased through an illustrative example. 
\end{abstract}

\section{Introduction}

The development of Model Predictive Control (MPC) \cite{Mayne} with stability and feasibility guarantees was initially focused on regulation around a fixed set-point. This was appropriate for most applications in the process industry \cite{Lee2011}, in which a set of optimal set-points is usually known from the plant design. Nevertheless, tracking changing set-points is also a necessity, not only in the process industry (e.g., in chemical batch processes \cite{ZoltanBratz03}), but also in many other applications such as HVAC systems \cite{AFRAM14} or mobile robot navigation \cite{ramirez1999nonlinear,Howard10}. However, feasibility might be lost under set-point changes in traditional MPC schemes.


An alternative solution for the problem of tracking under constraints are the so-called Reference and Command Governors \cite{GARONE17}. These constrained control schemes compute at each time instant an artificial reference to be applied to the system making use of online optimization. This artificial reference is such that, were it to be applied to the system, constraints would be fulfilled at all times. A different and novel approach, the Explicit Reference Governor \cite{nicotra2016constrained}, deals with constrained reference tracking without resorting to on-line optimization.

The addition of a decision variable resembling an artificial reference has also been used in MPC for linear systems \cite{rossiter1996}, as well as a time varying set-point that acts like a disturbance to be rejected \cite{pannocchia2005}. Subsequently, an MPC for Tracking for linear systems -- closely related to the one used in this paper -- was presented in \cite{LIMON2008}. This strategy solves the tracking problem by using an artificial reference for the steady state and inputs, a cost function that penalizes the deviation of the state from the artificial steady state reference, an additional term that penalizes the difference between the artificial reference and the actual set-point, and an extended terminal constraint based on an tracking invariant set.

An extension of the MPC for Tracking scheme able to deal with constrained nonlinear systems was presented in \cite{limon2018nonlinear}, where stability and feasibility were rigorously discussed. particularly the case in which the terminal constraint can be removed. However, in the presence of non-convex admissible output sets, this formulation might present convergence issues.

Although this limitation is not very stringent for some classical applications (\textit{e.g.} in process control), there are several cases in which state constraints are non-convex, such as mobile robot navigation \cite{Hagenars04}, formation flight control \cite{prodan2013}, aerospace problems like rendezvous, orbital transfer, optimal launch \cite{Liu14}, or soft landing maneuvers \cite{ACIKMESE2011341}. In the tracking scheme of \cite{limon2018nonlinear}, the way to deal with non-convex constraints is to restrict the operation of the MPC to a convex subset of admissible outputs. Although this practice can work for some applications, it introduces a relevant amount of conservativity.

To tackle this problem, in the preliminary paper \cite{cotorruelo2018tracking} a first extension to MPC for Tracking was presented which uses a homeomorphism to map a non-convex set of admissible outputs into a convex one. The use of this homeomorphism allows the MPC to deal with non-convex admissible output sets by solving the MPC problem in the convex domain. The main difficulty to do so is to find a suitable convexifying homeomorphism which, moreover, should be in closed form.

In this paper these preliminary results are further polished and more clearly stated. Furthermore, to approach the need for a homeomorphism in closed form we introduce a broad novel class of non-convex sets, the so called \textit{normal sets}, for which we provide a convexifying homeomorphism for this whole class. We subsequently modify the MPC for Tracking scheme so as to accommodate and compute this homeomorphism within the optimization problem, therefore solving the tracking problem in the case in which the admissible output set is a member of this family of non-convex sets.

This paper is organized as follows: in Section \ref{sec:ps} the problem is stated, followed by a brief summary of the MPC for Tracking formulation \cite{limon2018nonlinear}. In Section \ref{sec:ext} the formulation is extended to deal with non-convex sets of steady state output admissible sets. A novel class of non-convex sets is introduced in Section \ref{sec:normal}. The proposed extension of the MPC for Tracking  formulation is particularized for this class of sets in Section \ref{sec:star}, and it is applied to an illustrative example in Section \ref{sec:example}. The paper ends with the conclusions.

\subsection*{Notation}
A boldfaced variable $\textbf{u}$ denotes a sequence of values (\textit{i.e.}, $\{u(0),u(1),\ldots,u(N-1)\}$). $\Vert \cdot \Vert$ denotes the Euclidean norm, and $\Vert \cdot \Vert_P$ denotes the weighted Euclidean norm, \textit{i.e.}, $\Vert x \Vert_P=\sqrt{x^\top P x}$, with positive definite $P$. \ $I_n$ denotes the $n$-dimensional identity matrix. We denote the concatenation of two vectors $x$ and $u$ as $(x,u)=[x^\top u^\top]^\top$. Let a generic set $\mathcal{S}\subseteq \mathbb{R}^n$ and the subspace $X=\spanop\{e_1,\ldots,e_{m}\}$, $n>m$, where $e_i$ is the $i$-th vector of the canonical basis. We define the orthogonal projection of $\mathcal{S}$ onto  $X$ as $\proj_X(\mathcal{S})=\{x\in X\ |\ \exists y\in \mathbb{R}^{n-m} \textrm{ s.t. } (x,y)\in\mathcal{S}\}$. For a set $\mathcal{S}$, $\inter \mathcal{S}$ denotes the interior of $\mathcal{S}$, and $\partial\mathcal{S}$ denotes its boundary. A function $\alpha:\mathbb{R}_+\rightarrow\mathbb{R}_+$ is a $\mathcal{K}_\infty$ function if it is continuous, strictly increasing, unbounded from above, and  $\alpha(0)=0$. A bivariate function $V(x,y):\mathbb{R}^n\times\mathbb{R}^p\rightarrow\mathbb{R}$ is positive definite if $V(x,y)\geq \alpha (\|x\|)\ \forall\ (x,y)$ with $\alpha(\cdot)$ a $\mathcal{K}_\infty$ function.

\section{Problem statement}\label{sec:ps}

Consider a system described by a discrete time, nonlinear, time invariant model
\begin{equation}
\begin{aligned}
\label{eq:system}
x^+&=f(x,u)\\
y&=h(x,u),
\end{aligned}
\end{equation}
where $x\in\mathbb{R}^n$ is the system state, $x^+\in\mathbb{R}^n$ is the successor state, $u\in\mathbb{R}^m$ is the current control action, and $y\in\mathbb{R}^p$ is the controlled output of the system.
The system is subject to constraints in the form
\begin{equation}\label{eq:cons}
    (x,u)\in\mathcal{Z},
\end{equation}
where $\mathcal{Z}\subset\mathbb{R}^n\times\mathbb{R}^m$ is a closed set with nonempty interior.

The control objective is to steer the system output, $y$, to the desired output, $y_t$, while fulfilling the constraints at all times. A possible solution to this problem is MPC for Tracking\cite{limon2018nonlinear}, which allows to deal with the tracking problem by introducing an artificial reference, $y_s$, as an extra decision variable. At every time step, the system output will be steered toward $y_s$, while $y_s$ itself will move toward $y_t$.

For a given $y_s$, the steady state and input of system \eqref{eq:system} are such that
\begin{IEEEeqnarray}{L}
x_s=f(x_s,u_s),\IEEEyessubnumber\\
y_s=h(x_s,u_s).\IEEEyessubnumber
\end{IEEEeqnarray}

A usual practice in constrained control is to define  the following restricted set as a way to avoid equilibrium points with active constraints:
\begin{equation}
    \hat{\mathcal{Z}}=\{z:z+e\in\mathcal{Z},\,\forall |e| \leq\varepsilon\},
\end{equation}
with an arbitrarily small $\varepsilon>0$.
Accordingly, the set of admissible steady states can be defined as
\begin{IEEEeqnarray}{L}
\mathcal{Z}_s=\{(x,u)\in \hat{\mathcal{Z}}\, :\, x=f(x,u) \},\\
\mathcal{Y}_s=\{y=h(x,u)\, :\, (x,u)\in\mathcal{Z}_s\}.
\end{IEEEeqnarray}
As it is usually the case in the MPC for Tracking literature \cite[Assumption 1]{limon2018nonlinear}, we assume that there exist locally Lipschitz functions $g_x:\mathcal{Y}_s\rightarrow\mathbb{R}^n$ and $g_u:\mathcal{Y}_s\rightarrow\mathbb{R}^m$ such that
\begin{equation}
    x_s=g_x(y_s),\,u_s=g_u(y_s).
\end{equation}
and moreover that an invariant set for tracking for system \eqref{eq:system} is known, whose definition we recall from  \cite{limon2018nonlinear}:
\begin{defn}\label{def:inv}
For a given set of constraints $\mathcal{Z}$, a set of admissible references $\mathcal{Y}_t\subseteq\mathcal{Y}_s$ and a local control law $u=\kappa(x,y_s)$, a set $\Gamma\subset\mathbb{R}^n\times\mathbb{R}^p$ is an (admissible) invariant set for tracking for system \eqref{eq:system} if for all $(x,y_s)\in\Gamma$, we have that $(x,\kappa(x,y_s))\in\mathcal{Z}$, $y_s\in\mathcal{Y}_t$, and $(f(x,\kappa(x,y_s)),y_s)\in\Gamma$.
\end{defn}

In \cite{limon2018nonlinear} a tracking control strategy for \eqref{eq:system} subject to \eqref{eq:cons} was presented as the solution to the following optimization problem
\begin{IEEEeqnarray}{lr}\label{eq:MPC4T}
\min_{\textbf{u},y_s} V_{N_c,N_p}(x,y_t;\textbf{u},y_s)\IEEEyessubnumber\label{eq:TMPC_first}\\
\nonumber \textrm{s.t.} \\
x(0)=x \IEEEyessubnumber\\
x(j+1)=f(x(j),u(j)),\quad j=0,\cdots,N_c-1 \IEEEyessubnumber\\
(x(j),u(j))\in\mathcal{Z},\quad j=0,\cdots,N_c-1 \IEEEyessubnumber\\
x(j+1)=f(x(j),\kappa(x(j),y_s))),\, j=N_c,\cdots,N_p-1 \IEEEyessubnumber\IEEEeqnarraynumspace\\
(x(j),\kappa(x(j),y_s))\in\mathcal{Z},\quad j=N_c,\cdots,N_p-1  \IEEEyessubnumber\\
y_s\in\mathcal{Y}_t\IEEEyessubnumber\\
(x(N_p),y_s)\in\Gamma \IEEEyessubnumber,\label{eq:TMPC_last}
\end{IEEEeqnarray}
where $\textbf{u}$ is the computed sequence of control actions, $N_c\leq N_p$ are the control and prediction horizon, respectively; $\kappa(x,y_s)$ is the terminal control law, and $\Gamma$ is an invariant set for tracking. The objective function of the optimization problem \eqref{eq:TMPC_first}-\eqref{eq:TMPC_last} is \begin{multline}
    V_{N_c,N_p}(x,y_t;\textbf{u},y_s)=\\
    \sum_{j=0}^{N_c-1}\ell(x(j)-g_x(y_s),u(j)-g_u(y_s))\\ \nonumber
    +\sum_{j=N_c}^{N_p-1}\ell(x(j)-g_x(y_s),\kappa(x(j),y_s)-g_u(y_s))\\
    +V_f(x(N_p)-g_x(y_s),y_s) \nonumber+V_O(y_s-y_t),
\end{multline}
where $\ell:\mathbb{R}^n\times\mathbb{R}^m\rightarrow\mathbb{R}$ is the stage cost function, $V_f:\mathbb{R}^n\times\mathbb{R}^p\rightarrow\mathbb{R}$ is the terminal cost function, and $V_O:\mathbb{R}^p\rightarrow\mathbb{R}$ is the offset cost function, all of them being positive definite functions.

Recalling from \cite{limon2018nonlinear}, the stage and offset cost functions, as well as as the set of feasible set-points must fulfill the following assumptions:
\begin{assumption}\label{assu:cost_fcn}
\begin{enumerate}
    \item There exists a $\mathcal{K}_\infty$ function $\alpha_\ell$ such that $\ell(z,v)\geq\alpha_\ell(|z|)$ for all $(z,v)\in\mathbb{R}^{n+m}$.
    \item The set of feasible set-points $\mathcal{Y}_t$ is a convex subset of $\mathcal{Y}_s$.
    \item The offset cost function $V_O:\mathbb{R}^p\rightarrow\mathbb{R}$ is a subdifferentiable convex positive definite function such that the minimizer     $$y_s^\ast=\arg \min_{y_s \in\mathcal{Y}_t} V_O(y_s - y_t)$$ is unique. Moreover, there exists a $\mathcal{K}_\infty$ function $\alpha_O$ such that $$V_O(y_s-y_t)-V_O(y_s^\ast-y_y)\geq\alpha_O(|y_s-y_s^\ast|).$$
\end{enumerate}
\end{assumption}

Additionally, in order for stability to be proven, the terminal set and cost must fulfill the following assumptions:
\begin{assumption}\label{assu:stability}
\begin{enumerate}
    \item $\Gamma$ is an invariant set for tracking for the system $x^+=f(x,\kappa(x,y_s))$.
    \item $\kappa(x,y_s)$ is a control law such that for all $(x,y_s)\in\Gamma$, the equilibrium point $x_s=g_x(y_s)$ and $u_s=g_u(y_s)$ is an asymptotically stable equilibrium point for the system $x^+=f(x,\kappa(x,y_s))$. Besides, $\kappa(x,y_s)$ is continuous at $(x_s,y_s)$ for all $y_s\in\mathcal{Y}_t$.
    \item $V_f(x-x_s,y_s)$ is a Lyapunov function for system $x^+=f(x,\kappa(x,y_s))$ such that for all $(x,y_s)\in\Gamma$ there exists constants $b>0$ and $\sigma>1$ which verify $$V_f(x-x_s,y_s)\leq b|x-x_s|^\sigma$$ and
    \begin{IEEEeqnarray*}{rCl}
    V_f(f(x,\kappa(x,y_s))-x_s,y_s)-V_f(x-x_s,y_s)\leq\\ -\ell(x-x_s,\kappa(x,y_s)-u_s).
    \end{IEEEeqnarray*}
    \end{enumerate}
\end{assumption}

\begin{theorem}\label{teorema}{\!\!\cite[Theorem 1]{limon2018nonlinear}}
Suppose that Assumption \ref{assu:stability} holds true, and consider a given constant set-point $y_t$. Then for any feasible initial state $x_0$, the system controlled by the MPC derived from the solution of \eqref{eq:MPC4T} is stable, fulfills the constraints throughout time, and converges to an equilibrium point such that:
\begin{enumerate}
    \item If $y_t\in\mathcal{Y}_t$, then $\lim_{k\rightarrow\infty}|y(k)-y_t|=0$.
    \item If $y_t\notin\mathcal{Y}_t$, then $\lim_{k\rightarrow\infty}|y(k)-y^\ast_s|=0$, where
    $$y_s^\ast=\arg \min_{y_s \in\mathcal{Y}_t} V_O(y_s - y_t).$$
\end{enumerate}
\end{theorem}

As it will be demonstrated through an example in Section \ref{sec:example}, convexity plays a substantial role in convergence. In fact, the convergence of this scheme was proved under the assumption of convexity of the offset cost function $V_O$ and of the set of feasible set-points $\mathcal{Y}_t$, which is ideally equal to the set of admissible set-points $\mathcal{Y}_s$. While $V_O$ can be chosen to be convex, $\mathcal{Y}_s$ depends on the system model and constraints to be considered. When $\mathcal{Y}_s$ is not convex, convergence can only be proved in a convex subset of feasible set-points $\mathcal{Y}_t \subseteq \mathcal{Y}_s$, which is added as constraint in the optimization problem.

The purpose of this work is to overcome this source of conservativity by presenting an extension to the MPC for Tracking that allows it to deal with non-convex set of admissible outputs.
\section{Proposed extension}\label{sec:ext}
A possible way to overcome the aforementioned limitation is to map the set of admissible outputs onto a convex set, and solve the MPC optimization problem in the convex space. To do this we require a \textit{homeomorphism} able to perform such a mapping.

\begin{defn}[Homeomorphism]
Let $\mathcal{X}\subset\mathbb{R}^n$ and $\mathcal{Y}\subset\mathbb{R}^n$ be two sets. A function $\phi:\mathcal{X}\rightarrow \mathcal{Y}$ is a homeomorphism if it is bijective, continuous, and its inverse function, $\phi^{-1}:\mathcal{Y}\rightarrow \mathcal{X}$, is continuous as well. If such a function exists, the sets $\mathcal{X}$ and $\mathcal{Y}$ are said to be \textit{homeomorphic}.
\end{defn}

\begin{assumption}\label{assu:exists_homeo}
There exists a Lipschitz continuous homeomorphism $\phi$ between $\mathcal{Y}_s$ and a convex set $\Theta$.
\end{assumption}

\begin{remark}\label{rmk:existence}
Such a homeomorphism exists if $\mathcal{Y}_s$ and $\Theta$ share the same dimension and genus\footnote{In topology, the genus of a surface is the largest number of non-intersecting simple closed curves that can be drawn on the surface without separating it. Roughly speaking, the genus of a set is its number of holes\cite{sieradski1992introduction}.}\cite{sieradski1992introduction}. Note that $\mathcal{Y}_s$ is a property of the system and might not fulfill the conditions of Assumption \ref{assu:exists_homeo}. If such conditions are not fulfilled, one can choose a $\mathcal{Y}_t\subset\mathcal{Y}_s$ such that $\mathcal{Y}_t$ and $\Theta$ meet the aforementioned conditions.
\end{remark}

In order for the MPC to be able to deal with non convex sets of admissible steady-state outputs, the optimization problem needs to be adjusted to accommodate the homeomorphism. In particular, the optimization problem will be solved in terms of $y_s=\phi(\theta)$. This change of variables leads to modifications in its objective function and constraints.

Concerning the objective function, the homeomorphism needs to be accounted for in the offset and terminal costs, which become $V_O(\theta-\phi^{-1}(y_t))$ and $V_f(x(N_p)-g_x(\phi(\theta)),\theta)$, respectively. The offset cost function now penalizes the deviation between $\theta$ and the transformation of the desired set-point, $\phi^{-1}(y_t)$, whereas the terminal cost function is now in terms of $\theta$. Similarly, in this formulation, the terminal control law is defined in terms of $\theta$, $\kappa(x,\theta)$.

Gathering all of the above, the optimization problem becomes
\begin{IEEEeqnarray}{lr}\label{eq:MPC4T_homeo}
\min_{\textbf{u},\theta} V_{N_c,N_p}(x,y_t;\textbf{u},\theta)\IEEEyessubnumber\label{eq:MPC_first}\\
\nonumber \textrm{s.t.} \\
x(0)=x \IEEEyessubnumber\\
x(j+1)=f(x(j),u(j)),\quad j=0,\cdots,N_c-1 \IEEEyessubnumber\\
(x(j),u(j))\in\mathcal{Z},\quad j=0,\cdots,N_c-1 \IEEEyessubnumber\\
x(j+1)=f(x(j),\kappa(x(j),\theta)),\, j=N_c,\cdots,N_p-1 \IEEEyessubnumber\IEEEeqnarraynumspace\\
(x(j),\kappa(x(j),\theta))\in\mathcal{Z},\quad j=N_c,\cdots,N_p-1  \IEEEyessubnumber\\
\theta \in \Theta\IEEEyessubnumber \\
(x(N_p),\phi(\theta))\in\Gamma \IEEEyessubnumber,\label{eq:MPC_last}
\end{IEEEeqnarray}
with objective function
\begin{multline}\label{eq:objfun2}
    V_{N_c,N_p}(x,y_t;\textbf{u},\theta)=\\
    \sum_{j=0}^{N_c-1}\ell(x(j)-\hat{g}_x(\theta),u(j)-\hat{g}_u(\theta))\\
    +\sum_{j=N_c}^{N_p-1}\ell(x(j)-\hat{g}_x(\theta),\kappa(x(j),\theta)-\hat{g}_u(\theta))\\
    +V_f\left(x(N_p)-\hat{g}_x(\theta),\theta\right) +V_O\left(\theta-\phi^{-1}(y_t)\right).
\end{multline}
where $\hat{g}_x(\cdot)=g_x(\phi(\cdot))$ and $\hat{g}_u(\cdot)=g_u(\phi(\cdot))$

For what concerns the terminal cost and set, note that they are independent from the homeomorphism and must fulfill Assumption \ref{assu:stability}. More precisely, the terminal cost and set are computed for the original system since the differences between the regular MPC for Tracking and the proposed modification lie exclusively in the optimization problem.

In this formulation, the offset cost function penalizes the deviation between $\theta$ and $\phi^{-1}(y_t)$, and therefore the point toward which the system will be driven is no longer the closest admissible point to $y_t$ in the original space. Rather, it is the closest admissible point to $\phi^{-1}(y_t)$ in the transformed space, $\tilde{y}_s^\ast$:
\begin{equation*}
\tilde{y}_s^\ast=\phi(\arg\min_{\theta_s\in\Theta} V_O(\theta_s - \phi^{-1}(y_t))).
\end{equation*}

Note that the MPC strategy obtained from \eqref{eq:MPC4T_homeo} is such that Assumptions \ref{assu:cost_fcn} and \ref{assu:stability} are fulfilled in the transformed space, and therefore maintains all the theoretical guarantees of the original one, such as closed loop stability and recursive feasibility.

Clearly, the most critical point is finding a suitable homeomorphism. Theoretically, if the conditions in Remark \ref{rmk:existence} are met, the existence of such a function is ensured. However, to the best of our knowledge, there are no systematic procedures to build these mappings in the literature except for some simple cases (\textit{e.g.} star-shaped sets \cite{cotorruelo2018tracking,nicotra2016constrained}). To this avail, in the following section we provide a novel homeomorphism that convexifies a large class of non-convex sets.
\section{The normal form}\label{sec:normal}
\begin{figure}[t]
    \centering
    \begin{tikzpicture}[B/.style = {decorate, decoration={brace,amplitude=5pt,raise=1pt,mirror}}, scale=1.6]
\draw (-.25,0) -- (3.25,0) -- (3,2) -- (2.2,2) -- (1.95,1) -- (1.05,1) -- (.8,2) -- (0,2) node[below right]{$\mathcal{S}$} -- cycle;
\draw[-latex] (.95,0) -- (.95,1.4);
\draw[B]  -- (.95,0) node[above=32pt,right=-35pt] {$\overline{f}(y')$} (.95,1.4);
\draw[fill] (.95,0) circle [radius=0.05] node [below] {$y'$};
\draw[ultra thick,dotted] (-1,0)--(4,0) node[right]{$Y_c$};
\draw[ultra thick] (-.25,0)--(3.25,0);
\end{tikzpicture}
    \caption{Example of a normal set $\mathcal{S}$, its basis $Y_c$ (dotted line), the projection of $\mathcal{S}$ onto $Y_c$, $\mathcal{S}^c$ (thick solid line), and a visualization of $\overline{f}(y')$, for a point $y'\in \mathcal{S}^c$. Note that, in this particular case, $\underline{f}(y)=0\ \forall y\in \mathcal{S}^c$}
    \label{fig:ex_dist}
\end{figure}
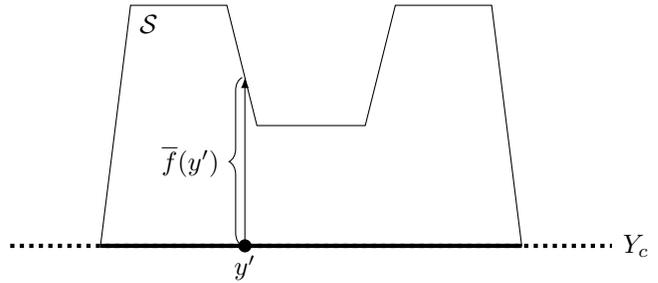

In this section we define a novel class of non-convex sets that can be transformed in a systematic way into a convex set. To do so, we first introduce the inverse image of the projection operator:
\begin{equation}\label{eqn:lifting}
    \proj^{-1}_{\mathcal{S},X}(q)=\{z\in \mathcal{S}\ |\ \proj_X(z)=q\}.
\end{equation}
\begin{defn}[Normal form]
Let $\mathcal{S}\subset\mathbb{R}^p$ be a bounded non-convex set described by a set of coordinates $y\in\mathbb{R}^p$. The set $\mathcal{S}$ is in \textit{normal form} if there exists a $(p-1)$-dimensional subspace $Y_c$ such that:
\begin{enumerate}
    \item $\mathcal{S}_c=\proj_{Y_c}(\mathcal{S})$ is convex,
    \item $\forall q \in \mathcal{S}_c$, $\proj^{-1}_{\mathcal{S},Y_c}(q)$ is simply connected.
\end{enumerate}

Without loss of generality, we will assume that $y=(Y_c,\, y_p)$, \textit{i.e.}, that $\proj_{Y_c}((Y_c,\,y_p))=Y_c$. For a set in normal form, $Y_c$ is called the \textit{basis} of $\mathcal{S}$. Additionally, if a set can be expressed in normal form with a basis $Y_c$, it is said to be \textit{normal} in $Y_c$. A depiction of an example of a normal set and its defining elements are shown in Figure \ref{fig:ex_dist}.
\end{defn}

In order to systematically construct a homeomorphism that maps a normal set $\mathcal{S}$ onto a convex one, we can take advantage of the fact that $\mathcal{S}$ is convex in the first $p-1$ coordinates (\textit{i.e.}, the basis), that is, the mapping needs only to modify the $p$-th coordinate. One possible way to achieve this is by means of a normalization. To do so, we introduce the functions $\overline{f},\,\underline{f}:\mathcal{S}_c\rightarrow\mathbb{R}$:
\begin{IEEEeqnarray*}{lCr}
\overline{f}(y)=\sup \Vert y-z \Vert, z \in \proj^{-1}_{\mathcal{S},Y_c}(y),\\
\underline{f}(y)=\inf \Vert y-z \Vert, z \in \proj^{-1}_{\mathcal{S},Y_c}(y).
\end{IEEEeqnarray*}
In the sequel, for notational simplicity, we will denote $\proj_{Y_c}(\cdot)$ and $\proj_{y_p}(\cdot)$ as $\cdot|_c$ and $\cdot|_p$, respectively. Gathering all of the above, the following function maps the interior of the normal set $\mathcal{S}$, $\inter\mathcal{S}$ onto a convexified version of itself, $\inter\mathcal{S}_c\times[0,1]$, where its $p$-th dimension has been normalized, removing all non-convexities.
\begin{equation}\label{eq:homeo}
    y=\phi(\theta)=\begin{bmatrix}
    \theta|_c\\
    \theta|_p(\overline{f}(\theta|_c)-\underline{f}(\theta|_c))+\underline{f}(\theta|_c)
    \end{bmatrix}.
\end{equation}
Conversely, its inverse function is
\begin{equation}\label{eq:homeoinv}
    \theta=\phi^{-1}(y)=\begin{bmatrix}
    y|_c\\
    \qfrac*{\displaystyle y|_p-\underline{f}(y|_c)}{\displaystyle \overline{f}(y|_c)-\underline{f}(y|_c)}
    \end{bmatrix}.
\end{equation}
Whenever $\overline{f}(\cdot)$ and $\underline{f}(\cdot)$ are continuous, \eqref{eq:homeo} and \eqref{eq:homeoinv} are continuous as well. Therefore \eqref{eq:homeo} is an homeomorphism by definition, as required in optimization problem \eqref{eq:MPC4T_homeo}.

\begin{remark}\label{rem:star}
Star-shaped sets are a particular class of normal sets. To see this, it is enough to realize that, when a $p$-dimensional star-shaped set is expressed in polar coordinates, (i) its first $(p-1)$ coordinates form a convex set, (ii) and the line segment joining the origin and any point in the boundary of the star-shaped set lies within the star-shaped set. Note that (i) and (ii) fall in the definition of a normal set, and that, for any star-shaped set $\mathcal{S}$, $\underline{f}(y)=0\ \forall y \in \mathcal{S}$.
\end{remark}
\section{Normal sets in Model Predictive Control}\label{sec:star}
In the case in which $\mathcal{Y}_s$ is normal in $Y_c$, we can apply \eqref{eq:homeo} to the MPC optimization problem \eqref{eq:MPC_first}--\eqref{eq:MPC_last}. However, the main drawback of this approach is that the characterization of $\overline{f}(\cdot)$ and $\underline{f}(\cdot)$ might prove cumbersome, especially in cases with large dimensionality. To tackle this, we will implicitly include these functions in the optimization problem so that the knowledge of the homeomorphism in closed form is not required. For this purpose, we make the following assumption:
\begin{assumption}
 There exists a function $\psi:\mathbb{R}^p\rightarrow\mathbb{R}$ such that\begin{equation}\label{eq:psi}
\begin{aligned}
\mathcal{Y}_s&=\{y\, :\, \psi(y)\geq0\}\\
\partial\mathcal{Y}_s&=\{y\, :\, \psi(y)=0\}.
\end{aligned}
\end{equation}
\end{assumption}

\begin{remark}
This assumption is not restrictive since we can use Zenkin's results \cite{zenkin1970analytic} to construct complex analytic shapes from the union and intersection of simpler ones, thus obtaining an arbitrarily good approximation of $\mathcal{Y}_s$.
\end{remark}

Following \eqref{eq:psi}, we can write $\overline{f}(y)$ and $\underline{f}(y)$ as follows.
\begin{equation}\label{eq:fbar}
\begin{aligned}
\overline{f}(y)&=\sup \overline{\lambda}\ : \ \psi\left(\begin{bmatrix}y|_c\\
\overline{\lambda}\end{bmatrix}\right)\geq0\\
\underline{f}(y)&=\inf \underline{\lambda}\ :\ \psi\left(\begin{bmatrix}y|_c\\
\underline{\lambda}\end{bmatrix}\right)\geq0.
\end{aligned}
\end{equation}

Instead of explicitly computing these functions, the values of $\overline{\lambda}$ and $\underline{\lambda}$ are included as decision variables in the MPC optimization problem. Thus, the resulting MPC optimization problem is as follows:
%
\begin{IEEEeqnarray}{lr}
 \min_{\textbf{u},\theta,\overline{\lambda},\underline{\lambda}} V_{N_c,N_p}(x,y_t;\textbf{u},\theta,\overline{\lambda},\underline{\lambda})\IEEEyesnumber \IEEEyessubnumber\label{eq:MPC4T_normal_first}\\
\nonumber \textrm{s.t.} \\
 x(0)=x,\IEEEyessubnumber\\
 x(j+1)=f(x(j),u(j)),\, j=0,\cdots,N_c-1\IEEEyessubnumber\\
 (x(j),u(j))\in \mathcal{Z},\, j=0,\cdots,N_c-1\IEEEyessubnumber\\
 x(j+1)=f(x(j),\kappa(x(j),\theta)),\, j=N_c,\cdots,N_p-1\IEEEeqnarraynumspace\IEEEyessubnumber\\
 (x(j),\kappa(x(j),\theta)\in \mathcal{Z},\  j=N_c,\cdots,N_p-1 \IEEEeqnarraynumspace\IEEEyessubnumber\\
\theta\in\Theta\IEEEyessubnumber\label{eq:mpc_theta}\\
(x(N_p),\phi(\theta))\in\Gamma \IEEEyessubnumber\\
 \psi\left(\begin{bmatrix}\theta|_c\\\overline{\lambda}\end{bmatrix}\right)\geq0\IEEEyessubnumber\label{eq:mpc_l1}\\
 \psi\left(\begin{bmatrix}\theta|_c\\\underline{\lambda}\end{bmatrix}\right)\geq0\IEEEyessubnumber\label{eq:MPC4T_normal_last}\\
\overline{\lambda}\geq \underline{\lambda} + \varepsilon,\IEEEyessubnumber\label{eq:mpc_last}
\end{IEEEeqnarray}
%
with $\varepsilon>0$  and $V_{N_c,N_p}(x,y_t;\textbf{u},\theta,\overline{\lambda},\underline{\lambda})$ is as in \eqref{eq:objfun2}, where $\phi(\cdot)$ is as in \eqref{eq:homeo}. Note that the term $\phi^{-1}(y_t)$ must be computed \textit{a priori} through \eqref{eq:homeoinv}.

The results of Theorem \ref{teorema} are still valid in this formulation, the proof is included in the Appendix.

\section{Illustrative example}\label{sec:example}

In order to illustrate the properties of the proposed methodology, we will apply it to a ball-on-plate system\cite{cagienard2007move}, which is a nonlinear positioning system widely used as benchmark for predictive schemes. In this case, we consider that the plate is non convex, which leads to a non-convex set of admissible equilibrium points.

We first demonstrate that the lack of convexity leads to the system controlled by a standard MPC for Tracking to get stuck at a certain point without converging to the target. Then we show that by using the proposed methodology, the derived controller can cope with this problem ensuring the convergence to the desired set-point.

The equations governing the evolution of the system are 

\begin{equation}\label{eq:sim_system}
\begin{aligned}
\Ddot{x}_1&=\frac{5}{7}(x_1\dot{\varphi}^2_1 + x_2\dot{\varphi}_1\dot{\varphi}_2 + g\sin{\varphi_1})\\
\Ddot{x}_2&=\frac{5}{7}(x_2\dot{\varphi}^2_2 + x_1\dot{\varphi}_1\dot{\varphi}_2 + g\sin{\varphi_2}),
\end{aligned}
\end{equation}
where $x_1$ and $x_2$ are the $x$ and $y$ position of the ball on the plate, respectively, $\varphi_1$ and $\varphi_2$ are the plate angles and $g$ is the gravitational acceleration. These equations were discretized using an Euler approach with a sampling time of $T_s=0.25\ \textrm{s}$.

The control input is the force exerted on the plate, $u=[\Ddot{\varphi}_1\,\Ddot{\varphi}_2]^\top$ and the output of the system is the position of the ball $y=[x_1\,x_2]^\top$. The system is subject to the following constraints:
\begin{equation}\label{eq:simsys_const}
\begin{aligned}
|u| \leq
\begin{bmatrix}
0.1\\0.1
\end{bmatrix}\\
y\in\mathcal{Y},
\end{aligned}
\end{equation}
where $\mathcal{Y}$ is the following set:
\begin{equation*}
\mathcal{Y}_s=\mathcal{E}_1\cup\mathcal{E}_2.
\end{equation*}
$\mathcal{E}_1$ and $\mathcal{E}_2$ are two ellipsoids, defined by their implicit equations
\begin{equation*}
\begin{aligned}
\mathcal{E}_1&=\{y:(y-y_{c_1})^\top P_1(y-y_{c_1})\leq 1\},\\
\mathcal{E}_2&=\{y:(y-y_{c_2})^\top P_2(y-y_{c_2})\leq 1\},
\end{aligned}
\end{equation*}
with parameters
\begin{IEEEeqnarray*}{lCr}
P_1=\begin{bmatrix}
16 & 0\\
0 & 0.5
\end{bmatrix},\,
P_2=\begin{bmatrix}
5.8551 & 7.3707\\
7.3707 & 10.6449
\end{bmatrix},\\
y_{c_1}=y_{c_2}=\begin{bmatrix}0 & 0\end{bmatrix}^\top.
\end{IEEEeqnarray*}
\begin{figure}[t]
    \centering
    \includegraphics[width=\linewidth]{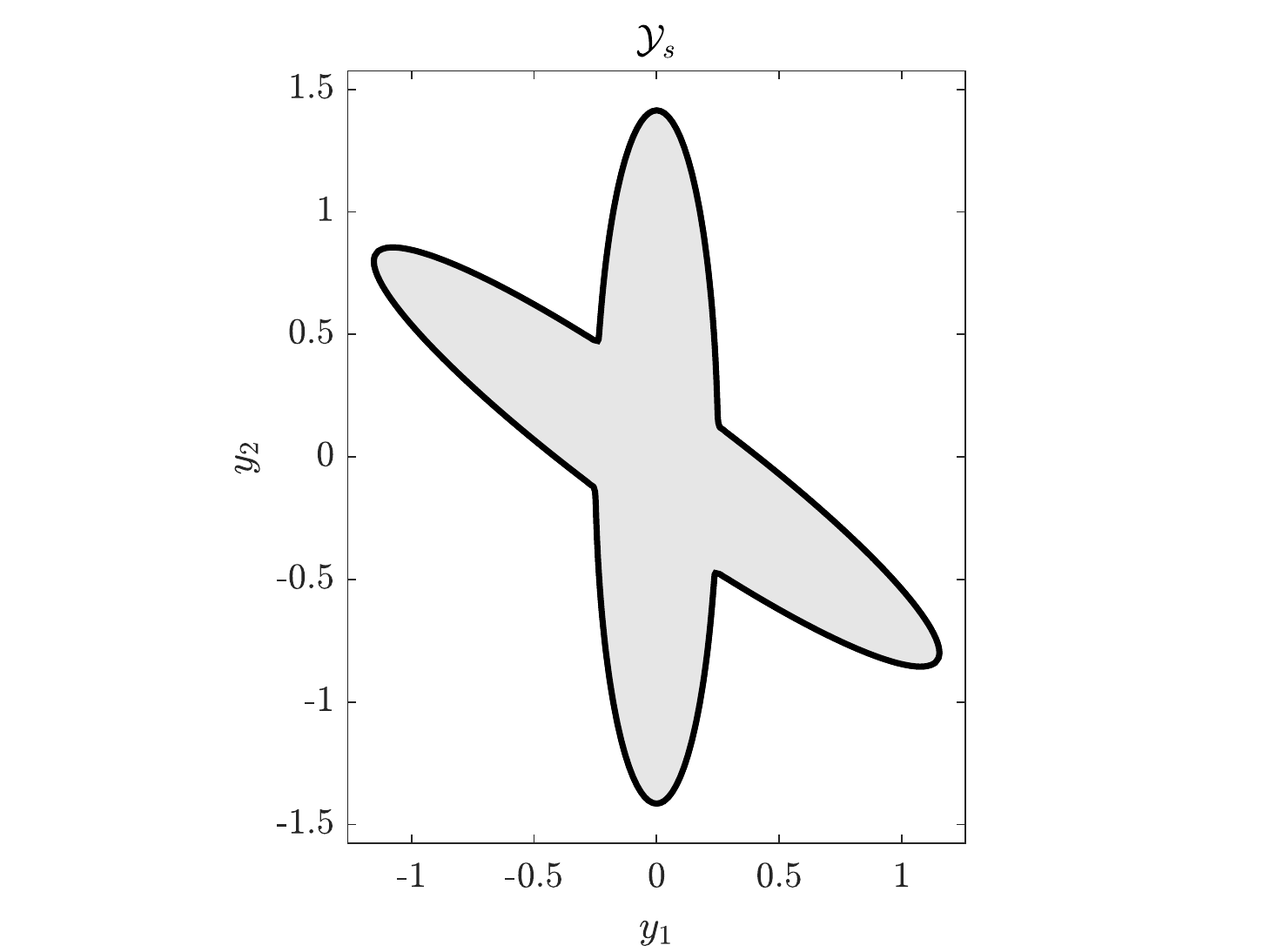}
    \caption{Set $\mathcal{Y}$}
    \label{fig:ys}
\end{figure}
A depiction of this set can be seen in Figure \ref{fig:ys}. Note that, in this case $\mathcal{Y}_s=\mathcal{Y}$.

\begin{figure}
    \centering
    \includegraphics[width=\linewidth]{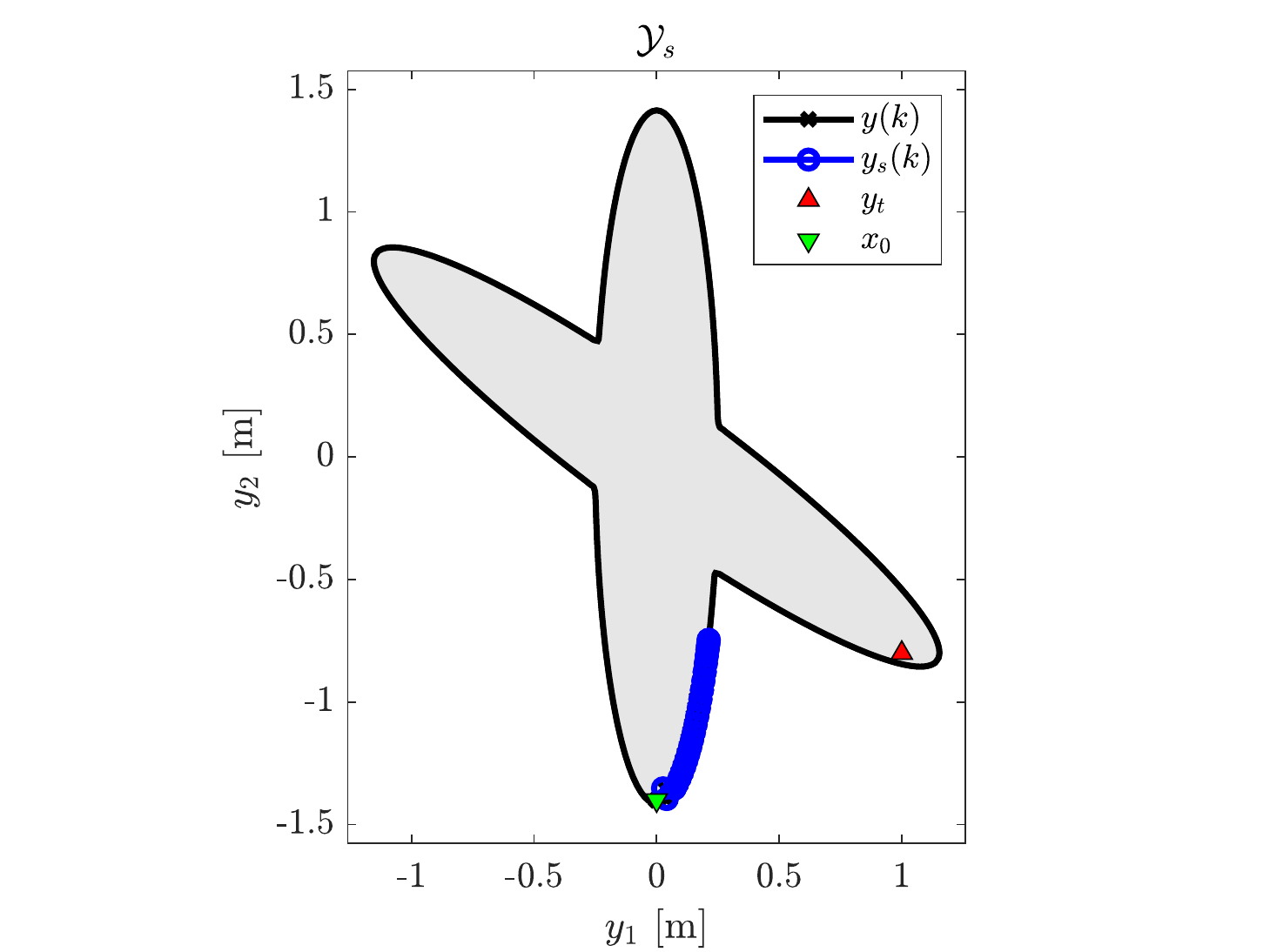}
    \caption{Evolution of the artificial reference, $y_s$ (solid black line with cross-shaped markers), and the output ,$y$ (solid blue line with circular markers), in the original space when the system is controlled with a traditional MPC for Tracking. The initial output of the system is depicted in a green downward pointing triangle, and the desired output, $y_t$, is depicted as a red upward pointing triangle.}
    \label{fig:output_nonconvex_no_diff}
\end{figure}

We used Zenkin's formulas to obtain the analytical description of $\mathcal{Y}_t$. For what concerns the terminal cost and set, we used the terminal equality constraint (see \cite[Sec.~3.A]{limon2018nonlinear}) with prediction and control horizons $N_p=N_c=4$. For the simulation, we set the initial conditions as $x_0=\begin{bmatrix}-0.1 & 0 & 1 & 0\end{bmatrix}^\top$ and the reference as $y_t=\begin{bmatrix}1&-0.8\end{bmatrix}^\top$. For what concerns the objective function, we used the following stage and offset cost for the standard MPC for Tracking:
\begin{equation}\label{eq:cost_function}
\begin{aligned}
\ell(x-x_s,u-u_s)&=\Vert x(j)-x_s \Vert^2_Q + \Vert u-u_s \Vert^2_R\\
V_O(y_s-y_t)&=\Vert y_s - y_t\Vert^2_T,
\end{aligned}
\end{equation}
with weighting matrices $Q=I_8$, $R=10I_2$, and $T=10^5I_2$. Simulation results for the standard MPC for Tracking are depicted in Figure \ref{fig:output_nonconvex_no_diff}. As it can be seen, the MPC for Tracking is not able to overcome the non convexities of $\mathcal{Y}_s$, therefore getting stuck and not reaching the desired set-point $y_t$.

We will now apply the proposed methodology to \eqref{eq:sim_system} subject to \eqref{eq:simsys_const}. Note that $\mathcal{Y}_s$ is a star-shaped set \cite{klain1997invariant}, which as previously demonstrated in Remark \ref{rem:star}, is a particular instance of normal sets.

For this case, the following cost functions were used:
\begin{IEEEeqnarray*}{lCr}
\ell(x-\hat{g}_x(\theta),u-\hat{g}_u(\theta)=\\
\Vert x-\hat{g}_x(\theta) \Vert^2_Q + \Vert u-\hat{g}_u(\theta) \Vert^2_R\\
V_O(\theta-\phi^{-1}(y_t))=\Vert \theta- \phi^{-1}(y_t)\Vert^2_T,
\end{IEEEeqnarray*}
where the parameters are the same as in \eqref{eq:cost_function}.
\begin{figure}[t]
    \centering
    \includegraphics[width=\linewidth]{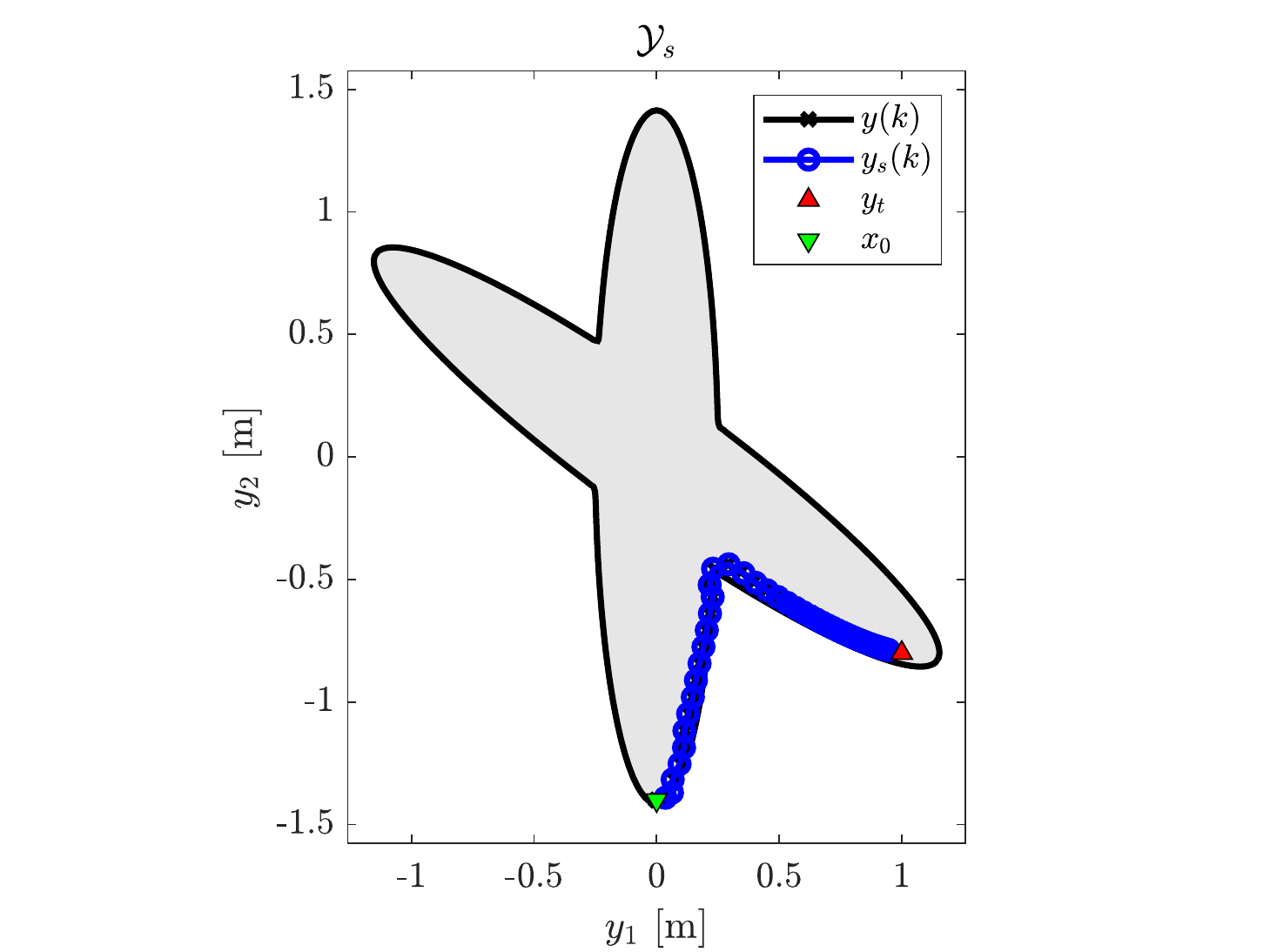}
    \caption{Evolution of the artificial reference, $y_s$ (solid black line with cross-shaped markers), and the output $y$ (solid blue line with circular markers), in the original space. The initial output of the system is depicted in a green downward pointing triangle, and the desired output, $y_t$, is depicted as a red upward pointing triangle.}
    \label{fig:output_ncvx}
\end{figure}
\begin{figure}[t]
    \centering
    \includegraphics[width=\linewidth]{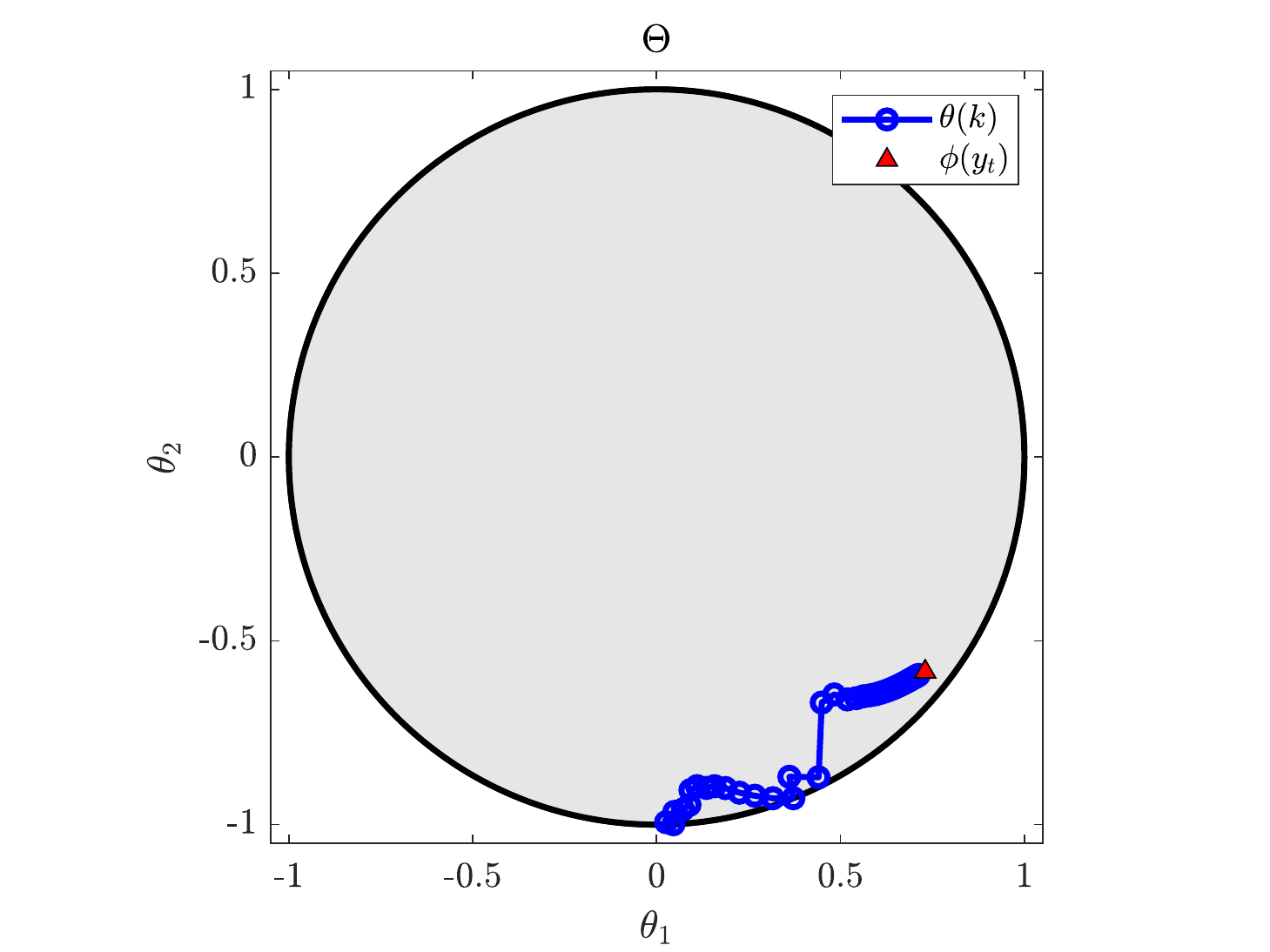}
    \caption{Evolution of $\theta$ (solid blue line with circular markers) in the transformed space. The transformation of the desired output, $\phi(y_t)$, is depicted as a red upward pointing triangle.}
    \label{fig:output_cvx}
\end{figure}
\begin{figure}[t]
    \centering
    \includegraphics[width=\linewidth]{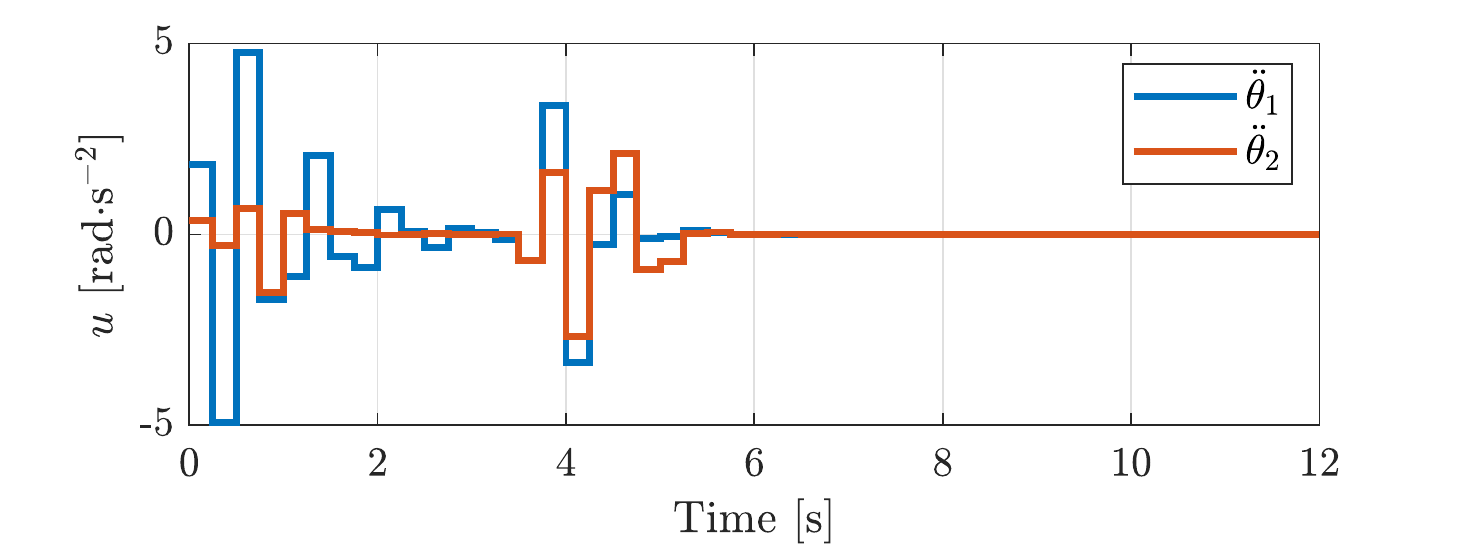}
    \caption{Time evolution of the control action.}
    \label{fig:control_action}
\end{figure}
Figures \ref{fig:output_ncvx}, \ref{fig:output_cvx}, and \ref{fig:control_action} depict the evolution of the output in the original space, the output in the transformed space, and the control action computed by the MPC, respectively. As it can be seen, this MPC formulation is able to drive the output to the desired set-point while fulfilling the constraints at all times.

To test the computational complexity of the proposed scheme, we measured the computational time of the MPC for Tracking with homeomorphism and compared it to a benchmark simulation without homeomorphism. For both cases, all the parameters and initial conditions were set equal. We obtained these computational times by measuring the average time taken by the optimizer over the whole simulation. 

These results can be found in Table \ref{tab:runtimes}, and, as it can be seen, the computational times of the simulation with the proposed extension are virtually the same as those of the benchmark test, which means that our scheme allows to deal with non-convex $\mathcal{Y}_s$ without increasing the computational complexity of the optimization problem with respect to the original scheme.

\begin{table}[ht]
\begin{center}
\begin{tabular}{|c||c|c|}
\hline
\textbf{Homeomorphism}& \textbf{Average time} & \textbf{Standard deviation} \\
\hline
Yes&0.1024 s & 0.0234 s\\
\hline
No & 0.1016 s & 0.0716 s\\
\hline
\end{tabular}
\end{center}
\caption{Average and standard deviation of the computational times of the simulation of the example, with and without the proposed extension.}
\label{tab:runtimes}
\end{table}

\section{Conclusions}\label{sec:conclusions}
In this paper, we presented an extension to the MPC for Tracking for non-convex admissible outputs sets. We defined a broad class of non-convex sets for which a convexifying homeomorphism is provided. We subsequently embedded this homeomorphism in the MPC for Tracking optimization problem. This embedding allows the homeomorphism to be computed within the optimization problem without the need for the homeomorphism closed form. The effectiveness of the proposed modification is shown in an illustrative example. Possible applications of this research may include robot navigation, obstacle avoidance, and UAV trajectory planning.

\bibliographystyle{IEEEtran}
\bibliography{thebib.bib}{}

\appendix[Proof of Theorem 1]

The proof of this theorem is akin to \cite[Theorem 1]{limon2018nonlinear}. The sole part of the proof that differs from it is Lemma 1, which is hereinafter included. The rest of the stability proof is identical to \cite[Theorem 1]{limon2018nonlinear}.

Consider system \eqref{eq:system} subject to \eqref{eq:cons} and assume that Assumptions 1, 2 and 3 hold. Consider a setpoint $y_t$ and assume that for a given state $x$ the optimal solution to \eqref{eq:MPC4T_normal_first}--\eqref{eq:MPC4T_normal_last} is such that $x=x^0_s(x,y_t,\theta,\overline{\lambda},\underline{\lambda})=\hat{g}_x(\theta(x,y_t))$. Then $V^0_{N_c,N_p}(x,y_t)=V_O(\theta^*-\phi^{-1}(y_t))$.

\textbf{PROOF:}

Consider that the optimal solution to \eqref{eq:MPC4T_normal_first}--\eqref{eq:MPC4T_normal_last} is $(\theta^0,\overline{\lambda}^0,\underline{\lambda}^0)$. Since $x=x_s^0$, the optimal value cost function is
\begin{equation*}
    V^0_{N_c,N_p}(x,y_t)=V_O\left(\theta^0-\phi^{-1}(y_t)\right).
\end{equation*}
As in \cite{limon2018nonlinear}, the lemma will be proved by contradiction; let us assume that 
$V_O(\theta^*-\phi^{-1}(y_t))>V_O(\theta^0-\phi^{-1}(y_t))$, then since $V_O$ is convex $\theta^0\neq \theta^*$. We now define $\hat{\theta}$ as
\begin{equation*}
    \hat{\theta}=\beta \theta^0 + (1-\beta)\theta^*,\, \beta \in [0,1].
\end{equation*}
Since $(\hat{g}_x(\theta),\hat{g}_u(\theta))\in\hat{\mathcal{Z}}$, there exists a $\hat{\beta}\in[0, 1)$ such that for a $\hat{\theta}$ with $\beta\in [\hat{\beta}, 1]$, the sequence of inputs generated by the terminal control law $\hat{\mathbf{u}}$ is such that $(\hat{\textbf{u}},\hat{\theta})$ is a feasible solution of \eqref{eq:MPC_first}--\eqref{eq:MPC_last}. Then, since using the extreme values $\overline{f}(\phi(\theta))$ and $\underline{f}(\phi(\theta))$ instead of $\overline{\lambda}$ and $\underline{\lambda}$ respectively in $V_{N_c,N_p}$ yields a suboptimal cost,
%
the following then holds
\begin{equation*}
    \begin{aligned}
    V_O\left(\theta^0\!-\!\phi^{-1}(y_t)\!\right)\!&=\!V_{N_c,N_p}^0(x_s^0,y_t)\\
    &\leq\!V_{N_c,N_p}\!\left(x_s^0,y_t;\!\textbf{u}^*\!,\theta,\overline{f}\left(\phi\left(\theta\right)\right)\!,\underline{f}\left(\phi\left(\theta\right)\right)\right)\\
    &\leq V_{N_c,N_p}\left(\!x_s^0,y_t;\!\hat{\textbf{u}},\hat{\theta},\overline{f}\left(\phi\left(\theta\right)\right)\!,\underline{f}\left(\phi\left(\theta\right)\right)\!\right)
    \end{aligned}
    \end{equation*}
where $\textbf{u}^*$ is the optimal solution to \eqref{eq:MPC_first}--\eqref{eq:MPC_last}. Since the last term of the previous inequality is equal to
\begin{multline*}
\sum_{j=0}^{N_p-1} \ell \left ( x(j) - \hat{g}_x (\hat{\theta}) , \kappa(x(j),\hat{\theta}) - \hat{g}_u (\hat{\theta}) \right)\\
+V_f \left( x(N_p) - \hat{g}_x (\hat{\theta}),\hat{\theta}\right)
+V_O\left(\hat{\theta}-\phi^{-1}(y_t)\right),
\end{multline*}
and the stage cost function $\ell$ being positive definite, it holds that
    \begin{equation*}\begin{aligned}
    %
     V_O\left(\theta^0\!-\!\phi^{-1}(y_t)\!\right)\!&\leq\!V_f\left(x_s^0\!-\!\hat{g}_x(\hat{\theta}),\hat{\theta}\right)\!+\! V_O\left(\hat{\theta}\!-\!\phi^{-1}(\hat{y_t})\right)\\
    &\leq b|x_s^0-\hat{g}_x(\hat{\theta})|^\sigma + V_O\left(\hat{\theta}-\phi^{-1}(y_t)\right)\\
    &\leq b\left(L_{\hat{g}}|\theta^0-\hat{\theta}|\right)^\sigma+ V_O\left(\hat{\theta}-\phi^{-1}(y_t)\right),\\
    \end{aligned}
\end{equation*}
where $ L_{\hat{g}}$ is the Lipschitz constant of $\hat{g}_x$. Taking into account that
\begin{multline*}
    b \left( L_{\hat{g}} | \theta^0 - \hat{\theta} | \right) ^ \sigma + V_O \left( \hat{\theta} - \phi^{-1} (y_t) \right) =\\
    L_{\hat{g}}^\sigma b (1 - \beta) ^\sigma | \theta^0 - \theta^* | ^\sigma + V_O \left( \hat{\theta} - \phi^{-1}(y_t) \right),
\end{multline*}
it holds that
\begin{multline*}
    V_O\left(\theta^0-\phi^{-1}(y_t)\right)\leq\\
    L_{\hat{g}}^\sigma b (1-\beta) ^ \sigma | \theta^0 - \theta^* | ^\sigma + V_O \left( \hat{\theta} - \phi^{-1}(y_t)     \right),
\end{multline*}
Since $V_O$ is convex 
\begin{multline*}
    V_O(\hat{\theta}-\phi^{-1}(y_t))\leq\\ \beta V_O(\theta^0-\phi^{-1}(y_t))+(1-\beta)V_O(\theta^*-\phi^{-1}(y_t)),
\end{multline*}
hence
\begin{multline*}
    V_O( \theta^0 - \phi^{-1}(y_t) ) \leq
    L_{\hat{g}}^\sigma  b (1-\beta)^\sigma |\theta^0 - \theta^*|^\sigma +\\ \beta V_O(\theta^0-\phi^{-1}(y_t)) + (1-\beta) V_O (\theta^* - \phi^{-1}(y_t)),
\end{multline*}
which in turn means that
\begin{multline*}
        V_O\left(\theta^0-\phi^{-1}(y_t )\right)-V_O\left(\theta^*-\phi^{-1}(y_t)\right)\leq\\ L_{\hat{g}}^\sigma b(1-\beta)^{\sigma-1}|\theta^0-\theta^*|^\sigma.
\end{multline*}
Since $\sigma > 1$, taking the limit from the left yields
\begin{multline*}
        V_O\left(\theta^0-\phi^{-1}(y_t )\right)-V_O\left(\theta^*-\phi^{-1}(y_t)\right)\leq\\
        \lim_{\beta\rightarrow 1^-}L_{\hat{g}}^\sigma b(1-\beta)^{\sigma-1}|\theta^0-\theta^*|^\sigma=0,
\end{multline*}
which contradicts the initial assumption. $\hfill\blacksquare$

\end{document}